\documentclass[aps,prl,twocolumn,superscriptaddress,hyperref]{revtex4}

\usepackage{graphicx}
\usepackage{latexsym}
\usepackage{amsmath}
\usepackage{amssymb}
\usepackage{comment}

\newcommand{\vect}[1]{{\mathbf #1}}

\begin{document}


\title{Polarized Fermi condensates with unequal masses: Tuning the tricritical
point}

\author{M. M. Parish}
\email{mmp24@cam.ac.uk}
 \affiliation{Cavendish Laboratory, JJ Thomson Avenue, Cambridge,
 CB3 0HE, UK}

\author{F. M. Marchetti}
\email{fmm25@cam.ac.uk} \affiliation{Cavendish Laboratory, JJ
Thomson Avenue, Cambridge, CB3 0HE, UK}

\author{A. Lamacraft}
\affiliation{Rudolf Peierls Centre for Theoretical Physics, 1 Keble
Road, Oxford OX1 3NP, UK} 

\author{B. D. Simons}
\affiliation{Cavendish Laboratory, JJ Thomson Avenue, Cambridge, CB3
0HE, UK}

\date{\today}

\begin{abstract}
We consider a two-component atomic Fermi gas within a mean-field,
single-channel model, where both the mass and population of each
component are unequal. We show that the tricritical point at zero
temperature evolves smoothly from the BEC- to BCS-side of the
resonance as a function of mass ratio $r$. We find that the interior
gap state proposed by Liu and Wilczek is always unstable to phase
separation, while the breached pair state with one Fermi surface for
the excess fermions exhibits differences in its DoSs and pair
correlation functions depending on which side of the resonance it
lies. Finally, we show that, when $r\gtrsim 3.95$, the finite
temperature phase diagram of trapped gases at unitarity becomes
topologically distinct from the equal mass system.
\end{abstract}

\pacs{}



\maketitle


Recent advances in the ability to manipulate and control ultra-cold
atomic vapors have provided a unique experimental system in which to
explore pairing phenomena. Following the successful realization of
the crossover from the BCS
to the Bose-Einstein condensed phase~\cite{BCS-BECexpt},
attention has turned to the consideration of more exotic Fermi
superfluids. A subject that has attracted particular
interest is that of Fermi condensates with imbalanced spin
populations~\cite{imbalance_theory,imbalance_expt},
having relevance to
QCD and magnetized superconductors~\cite{casalbuoni2004}. Equally
intriguing is the case in which both the mass and population of each
fermionic species in the condensate are unequal. Indeed, the
realization of Feshbach resonances in Fermi-Bose
mixtures~\cite{unequal_expt}, 
and the predicted stability of diatomic molecules close to resonance
over a wide range of mass ratios~\cite{petrov2005}, suggest that
such superfluid mixtures should be experimentally accessible.

Previous
studies~\cite{liu2003,unequal_theory,wu2006,lin2006} have raised
several important issues unique to Fermi condensates with spin and
mass imbalances. Firstly, it has been proposed that the breached
pair
(BP)
state, where the superfluid and excess fermionic states phase
separate in momentum space, can possess excess fermions with two
Fermi surfaces (BP-2) ---
the interior gap
state~\cite{liu2003}.  In the case of equal masses, the
BP
state can only have one Fermi surface (BP-1). However, it is
still unclear whether BP-2 can become stable for large mass ratios
near the Feshbach resonance~\cite{forbes2005}, or whether it is
always unstable to phase separation in real space as in the weak
coupling limit~\cite{bedaque2003}. Secondly, although there have
been studies of the zero temperature phase diagram for the
homogeneous gas with unequal masses~\cite{unequal_theory,wu2006}, so
far no tricritical point (such as that discussed in the equal mass
system~\cite{parish2006,gubbels2006}) has been
identified. It is natural to ask whether such a tricritical point is
generic and, if so, how it evolves
with mass ratio.
Finally, it has been
shown that trapped Fermi gases with
unequal masses can exhibit spatial phase separation at zero
temperature that differs qualitatively from that of the equal mass
case~\cite{lin2006}. Whether and how such features extend to
finite temperatures
remains unanswered.
Focussing on the homogeneous
system, all three issues will be addressed in this
work. To focus our discussion, we will
address the phase
boundaries between spatially homogeneous phases, leaving
the potential for Fulde-Ferrell-Larkin-Ovchinnikov
phases~\cite{FFLO} to future investigation.

Referring to
wide (viz. entrance-channel dominated) Feshbach resonances~\cite{Koehler2006},
we restrict
attention to a single-channel Hamiltonian
with contact potential:
%
\begin{multline}\label{eq:ham}
  \hat{H} - \sum_{\sigma = \uparrow,\downarrow} \mu_\sigma
  \hat{n}_{\sigma}  = \sum_{\vect{k}\sigma}
  \left(\epsilon_{\vect{k} \sigma} - \mu_{\sigma}\right)
  c_{\vect{k} \sigma}^\dag c_{\vect{k} \sigma} \\
   + \frac{g}{V} \sum_{\vect{k},\vect{k}',\vect{q}}
  c_{\vect{k}+\vect{q}/2 \uparrow}^\dag c_{-\vect{k}+\vect{q}/2
  \downarrow}^\dag c_{-\vect{k}'+\vect{q}/2 \downarrow}
  c_{\vect{k}'+\vect{q}/2 \uparrow}\; .
\end{multline}
Here, $\epsilon_{\vect{k}\sigma}=\frac{{\vect{k}}^2}{2m_{\sigma}}$
($\hbar=1$), $g$ is the interaction strength, $V$ is the volume, and
we allow the mass $m_{\sigma}$ and chemical potential $\mu_{\sigma}$
of each spin to be different, with average chemical potential
$\mu=(\mu_{\uparrow}+\mu_{\downarrow})/2$ and `Zeeman' field
$h=(\mu_{\uparrow}-\mu_{\downarrow})/2$. To obtain
the topology of the phase diagram spanning the BCS-BEC limits for
unequal masses (including the locus of the finite-temperature
tricritical point at which the transition between the superfluid and
normal phases switches from first- to second-order), we will develop
a mean-field analysis of the system analogous to that presented for
equal masses~\cite{parish2006}. While such a treatment is not
expected to be quantitatively correct for all interaction strengths,
it should provide a
reliable qualitative description even close to unitarity.

Setting
$\epsilon_{\vect{k}+}=\frac{\vect{k}^2}{2m_r}$, $\frac{2}{m_r}=
\frac{1}{m_{\uparrow}}+\frac{1}{m_{\downarrow}}$, $\epsilon_{\vect{k}-}=
\epsilon_{\vect{k}+}\frac{(r-1)}{(r+1)}$, and $r=\frac{m_{\downarrow}}{
m_{\uparrow}}$,
%
the free energy density can be expressed as
\begin{multline}
  \Omega^{0} (\mu,h) = \min_{\Delta}\left\{-\frac{\Delta^2}{g}
  \phantom{\frac{1}{V}\sum_{\vect{k}}\left[\frac{1}{\beta }\right] }
  \right. \\ \left. + \frac{1}{V}\sum_{\vect{k}}\left[ \xi_{\vect{k}}
  - E_{\vect{k}} -\frac{1}{\beta }\sum_{\sigma} \ln\left(1+e^{-\beta
  E_{\vect{k\sigma}}}\right)\right]\right\}\; ,
\label{eq:energy}
\end{multline}
where $\min_{\Delta}$ gives the \emph{global} minimum with respect
to $\Delta$ (to be calculated numerically). Here, $\xi_{\vect{k}} =
\epsilon_{\vect{k} +} - \mu$ and, defining $E_{\vect{k}}=
\sqrt{\xi_{\vect{k}}^2 + \Delta^2}$, the quasiparticle energies are
given by $E_{\vect{k} \sigma}=E_{\vect{k}} \mp (h-
\epsilon_{\vect{k}-})$. The free energy~\eqref{eq:energy} differs
from that of the equal mass system only through the appearance of a
momentum-dependent contribution, $\epsilon_{\vect{k}-}$, to the
Zeeman term which leads to the symmetry $(h,r)\mapsto (-h,1/r)$.
We introduce the $s$-wave scattering length $a$ via the prescription
$\frac{m_rV}{4\pi a} = \frac{V}{g} + \sum_{\vect{k}}^{k_0}\frac{1}{
2\epsilon_{\vect{k} +}}$,
%
%
where the
UV cutoff $k_0$ can be sent to infinity at the
end of the calculation.
%
With experiments
performed at fixed
density, $\mu$
and $h$ are
determined from the total density $n \equiv
n_{\uparrow} + n_{\downarrow} = - \partial \Omega^0/\partial \mu$ and
the population imbalance 
$m \equiv n_{\uparrow} - n_{\downarrow} = -\partial
\Omega^0/\partial h$ \footnote{The definition of $m$ also leads to
the
symmetry $(m,r) \mapsto (-m,1/r)$,
allowing restriction
to $m\geq0$ when varying $r$.}.
Note that the stability criterion adopted in
Refs.~\cite{unequal_theory,wu2006} corresponds only to finding
\emph{local} minima with respect to $\Delta$ in
Eq.~\eqref{eq:energy} and thus will not correctly locate a first
order transition.  The presence of such a transition implies a
region of phase separation in real space if $n$ and $m$ are held
fixed~\cite{bedaque2003}. The two phases in question are a normal
phase with $\Delta=0$ and a superfluid phase with $\Delta\neq 0$,
which is less magnetized as pairing tends to enforce equality of
populations.

\begin{figure}
\centering
\includegraphics[width=0.4\textwidth]{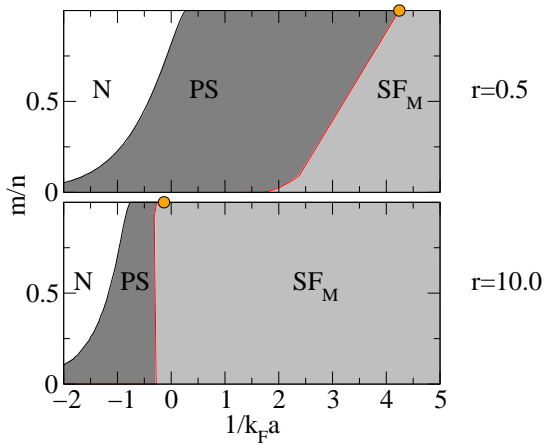}
\caption{(Color online) Zero temperature phase diagrams for different
         mass ratios $r=m_{\downarrow}/m_{\uparrow}$, where a positive
         polarization $m$ corresponds to an excess of particles with
         spin $\uparrow$. All mass ratios exhibit the same three
         phases: the normal state (N), the spatially homogeneous
         superfluid (SF$_{\rm M}$), and phase separation (PS) between normal
         and superfluid states in real space.
         The region of
         SF$_{\rm M}$ expands with increasing $r$. The boundaries enclosing the
         PS region are all first-order, while the tricritical point is
         represented by a circle. The line defined by $m/n=0$ corresponds to
         the usual BCS-BEC crossover.} \label{fig:zeroT}
\end{figure}
\begin{figure}
\centering
\includegraphics[width=0.3\textwidth]{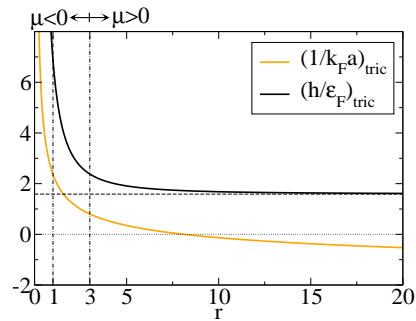}
\caption{(Color online) Evolution of the tricritical point
         $(1/k_Fa, h/\varepsilon_F)_{\text{tric}}$ as a function of
         mass ratio $r$
         with $h>0$. The values
         of $r=1$, the equal mass case, and $r=3$, where the chemical potential
         $\mu$ changes sign, are explicitly marked.} 
\label{fig:tricrit}
\end{figure}

Beginning with zero temperature
(Fig.~\ref{fig:zeroT}), we find that the basic structure of the phase
diagram ($\frac{1}{k_F a}$, $\frac{m}{n}$) for unequal masses mirrors that
of the equal mass system \footnote{We define the Fermi momentum with
respect to
average density $n/2$, so that $k_F=(3\pi^2n)^{1/3}$
and $\varepsilon_F=k_F^2/2m_r$.}. Phase separation (PS) is found
between the normal (N) and superfluid states. On the BCS side the
superfluid component is always unpolarized, while the
BP
or magnetized superfluid (SF$_{\rm M}$) state exists for
sufficiently strong interaction and eventually undergoes a
second-order transition to the fully-polarized normal state beyond
the tricritical point. However, the position of the tricritical
point is dramatically shifted, with the regions of PS and SF$_{\rm
M}$ shrinking and expanding, respectively, as $r$ increases
(in qualitative agreement with
Refs.~\cite{unequal_theory,wu2006,lin2006}).

We can gain further insight into the phase diagram 
by examining the behavior of the tricritical point as a function of
mass ratio, $r$. As shown in Fig.~\ref{fig:tricrit}, the tricritical
point, determined as the point at which both the quadratic and
quartic terms in the Landau expansion of the free
energy~\eqref{eq:energy} vanish, evolves smoothly from the BEC to
BCS limits, with $(h/\varepsilon_F,1/k_Fa)_{\text{tric}} \rightarrow
(\infty,\infty)$ as $r\rightarrow 0$, and $(h/\varepsilon_F,
1/k_Fa)_{\text{tric}}\rightarrow (2^{2/3},-\infty)$ as $r\rightarrow
\infty$. Moreover, as in the equal mass system, the tricritical
point always corresponds to a fully-polarized state, $m/n=1$
\footnote{This feature survives Nozi\`eres-Schmitt-Rink fluctuation
corrections, as in the equal mass case~\cite{parish2006}, but other
corrections that include interactions between excess fermions and
molecular bosons may modify it.}. It is interesting to note
(Fig.~\ref{fig:tricrit}) that the chemical potential, $\mu$, at the
tricritical point becomes positive for $r\ge 3$, hinting at the
possibility of a BP-2 state.
However, a mean-field analysis of the phase boundaries in
the limit $r\rightarrow\infty$ shows that only the BP-1 state is
ever stable. As in the weak-coupling limit~\cite{bedaque2003}, phase
separation is always favored over the BP-2
state~\footnote{The existence of a BP-2 phase for $r\ll 1$ was
discussed in Ref.~\cite{liu2003}, but the possibility of phase
separation was ignored.}.

\begin{figure}
\centering
\includegraphics[width=0.33\textwidth]{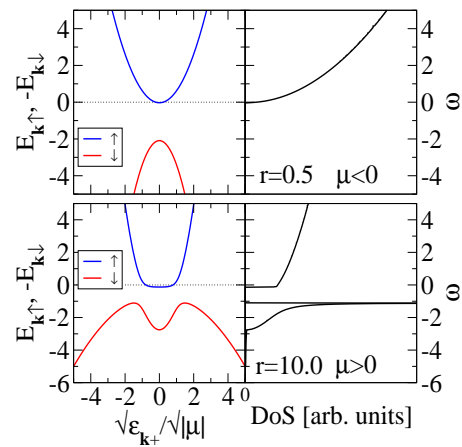}
\caption{(Color online) Quasiparticle dispersions $E_{\vect{k}
    \uparrow}$, $E_{\vect{k} \downarrow}$ together with the majority species
    ($\uparrow$)
    projection of the
    $\text{DoS} (\omega) =
    -\frac{1}{\pi} \sum_{\vect{k}}\Im [G_{11}(\vect{k},i\epsilon_n
    = \omega + i \eta)]$ for
    mass ratios
    $r=0.5$ (upper panels) ($\Delta/|\mu|=0.25$, $h/|\mu|=1.06$,
    $1/k_Fa=2.93$, $m/n=0.12$) and $r=10.0$ (lower panels)
    ($\Delta/\mu=0.85$, $h/|\mu|=1.44$, $1/k_Fa=-0.07$, $m/n=0.16$).}
\label{fig:dist1}
\end{figure}
\begin{figure}
\centering
\includegraphics[width=0.4\textwidth]{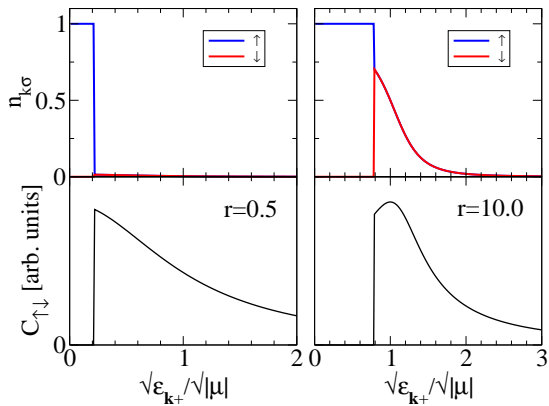}
\caption{(Color online) Averaged momentum distribution for each
    spin state $n_{\vect{k} \uparrow, \downarrow}$ and correlation
    function $C_{\uparrow \downarrow}$ for $r=0.5$ (left panels) and
    $r=10.0$ (right panels) and the same parameter values used in
    Fig.~\ref{fig:dist1}.}
\label{fig:dist2}
\end{figure}

Even though the
BP
state only ever possesses one Fermi
surface, the associated quasiparticle excitation spectrum can
display minima at \emph{non-zero} $\vect{k}$, a property which is
also characteristic of the BP-2 phase. As demonstrated in
Fig.~\ref{fig:dist1}, while the spectrum remains gapless for both
$r=0.5$ and $r=10$ (i.e., the density of states, DoS($\omega$), is
finite at $\omega=0$), the excitation spectrum for $r=10$ also
exhibits stationary points at non-zero $\vect{k}$ resulting in a
square-root singularity in the associated DoS. Such a singularity
may be regarded as `BCS-like', while its absence for $r=0.5$ is
`BEC-like' (cf. quasiparticle excitation spectra at the BCS-BEC
crossover in the equal mass case~\cite{parish2005_2}). More
generally, DoS singularities in the
BP
state only occur
for $r>1$, where the species with the smaller mass is in the
majority. When $\mu>0$, the singularities exist for all
polarizations, but when $\mu<0$, they are restricted to the region
around $m=0$ where $\Delta/|\mu| > 2\sqrt{r}/(r-1)$~\footnote{For
zero polarization, while it is true that the ground
state remains insensitive to mass ratio (as emphasized in
Ref.~\cite{wu2006}), the quasiparticle excitation spectrum and its
associated DoS are strongly dependent on it. In particular,
only one branch of the spectrum has a minimum at non-zero
$\vect{k}$ when $|\mu/\Delta| < |r-1|/2\sqrt{r}$, while both
branches have minima at non-zero $\vect{k}$ when $|r-1|/2
\sqrt{r}<\mu/\Delta$ and neither branch possesses them when
$\mu/\Delta <-|r-1|/2\sqrt{r}$.}. Such dramatic differences in the
DoS should be accessible experimentally using optical
excitations~\cite{torma2000}.

Further signatures of breached pairing are visible in the momentum
distribution, $n_{\vect{k}\sigma}$, and correlation function,
\begin{multline*}
  C_{\uparrow\downarrow}(\vect{k_1},\vect{k_2}) = \langle\hat{n}_{\vect{k_1}
\uparrow} \hat{n}_{\vect{k_2}\downarrow}\rangle -\langle
\hat{n}_{\vect{k_1}\uparrow} \rangle \langle\hat{n}_{\vect{k_2}
\downarrow}\rangle\\
  = \delta_{\vect{k_1},-\vect{k_2}} \frac{\Delta^2}{4E_{\vect{k_1}}^2}
  [1-\Theta(-E_{\vect{k} \uparrow})-\Theta(-E_{\vect{k}
  \downarrow})]^2\; .
\end{multline*}
Referring to Fig.~\ref{fig:dist2}, breached pairing is characterized
by a phase separation in momentum space between the excess of
majority species $\uparrow$ and the minority species $\downarrow$
involved in the superfluid state. In both cases, the correlation
function $C_{\uparrow\downarrow}$ shows a `hole' for momenta less
than the Fermi momentum of the majority quasiparticles. This hole at
small momenta is reminiscent of the Pauli blocking observed in the
closed-channel molecule of the $^{40}$K Feshbach
resonance~\cite{K40_FR} where there is always an inherent particle
number asymmetry in the open channel, even in the usual BCS-BEC
crossover. The correlation function $C_{\uparrow\downarrow}$ can
also directly probe the sign of $\mu$: for $\mu>0$ we have a peak
beyond the blocked region, as shown in the $r=10.0$ case, provided
$\Delta/\mu > |h/\mu + (1-r)/(1+r)|$.
In principle, such a feature can be detected experimentally using
noise correlations~\cite{altman2004}.

\begin{figure}
\centering \includegraphics[width=0.45\textwidth]{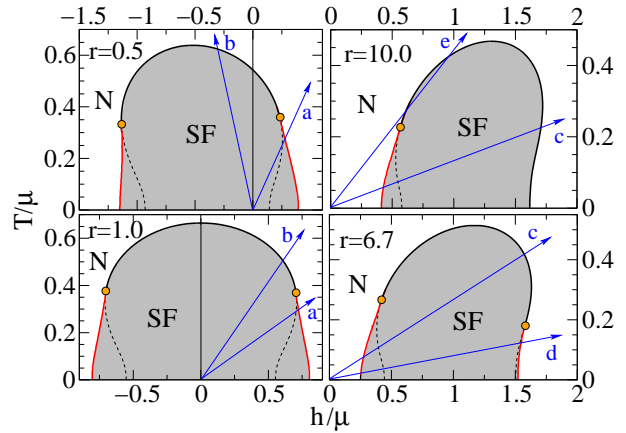}
\caption{(Color online)
         Phase diagrams in the
         $(T/\mu,h/\mu)$-plane for different mass ratios $r$ at unitarity
         $1/k_Fa = 0$. Mass ratio $r=6.7$ corresponds to the case of
         $^6$Li-$^{40}$K mixtures.
         The transition between the N and SF phase can
         be first (red) or second order (black solid), separated by
         the tricritical point. The dashed black line describes the
         second order transition in the region where the true transition is
         first order.
         Lines of constant gradient $T/h$ represent the trajectory
         traveled from the center to the edges of the trap, examples
         of which are shown by the arrows (a-e). The starting points on these
         lines are the values of $(T/\mu,h/\mu)$ at the trap center.}
\label{fig:Ttrap}
\end{figure}

Turning to finite temperature, the phase diagram for unequal masses
in the homogenous system is qualitatively similar to the equal mass
case, but the topology of phases within a trap can be different.
A study of $^6$Li-$^{40}$K mixtures at zero
temperature~\cite{lin2006} has already revealed that the unequal
mass case offers a richer variety of phase-separated states. In
addition to configurations where there is a superfluid core
surrounded by the normal state (as in the equal mass case), the
superfluid region can occupy a shell sandwiched between a normal
inner core and a normal outer shell with opposite polarizations. To
investigate the effects of temperature $T$ and different mass ratios
$r$ in the trap geometry, we restrict ourselves to the case in which
the local density approximation can be applied. Thus, assuming each
species experiences the same trapping potential $V(\vect{r})$, the
effects of the trap can be included in the spatially-varying chemical
potential, $\mu - V(\vect{r})$, while the term $h$ remains fixed.

In Fig.~\ref{fig:Ttrap}, we plot mean-field phase diagrams at
unitarity as a function of $T/\mu$ and $h/\mu$. Note that
Nozi\`eres-Schmitt-Rink fluctuations
will not alter these
diagrams since
corrections only enter into the density, $n$, and
the polarization, $m$~\cite{nozieres1985}. We restrict ourselves to
the case where $\mu>0$ because this is enough to completely
encompass the SF-N phase boundaries, even though a fully-polarized
normal state obviously exists for $-|h|<\mu<0$. While the equal mass
($r=1$) phase diagram is symmetric around $h=0$ as expected, we see
that lowering or increasing $r$ translates the superfluid dome to
the left or right, respectively. For sufficiently small
translations, we have the ordinary trapped case of a superfluid core
with a first (arrow $a$) or second (arrow $b$) order transition to
the surrounding normal state. However, once $r\gtrsim 3.95$, the
superfluid region has shifted entirely to the $h>0$ plane, as shown
for $r=6.7$ and $r=10$ \footnote{Equivalently, the whole
superfluid region will shift to the $h<0$ plane if $1/r\gtrsim
3.95$.}.  Provided $h/\mu$ is sufficiently small at the trap center,
this naturally leads to a superfluid shell structure
where the superfluid phase is sandwiched between a `heavy' normal
core and an outer `light' normal phase.
This structure can either have two first-order SF-N phase boundaries ($d$),
one first and one second ($c$), or two second-order phase boundaries
($e$), depending on the value of $T/h$. The case considered in
Ref.~\cite{lin2006} corresponds to the $T/\mu=0$ axis in the $r=6.7$
phase diagram, and is thus an example of category $d$. Note,
however, that one tricritical point vanishes when $r$ is further
increased (corresponding to the point where $(1/k_Fa)_{\text{tric}}$
changes sign in Fig.~\ref{fig:tricrit}), so that `$d$-type' SF
shells are then no longer possible.

In conclusion, we have investigated the mean-field phase diagram of
a polarized Fermi condensate with unequal masses.  We have shown
that the zero temperature tricritical point smoothly crosses over
from $1/k_Fa>0$ to $1/k_Fa<0$ as $r$ increases, but the interior gap
state is never stable, even in the limit of infinite $r$. However,
differences in the
BP
states do show up in the DoS and
pair correlations. Finally, we show how the phase diagram of trapped
gases depends on $r$ and $T$, including how one obtains superfluid
shells for $r\gtrsim 3.95$.





\end{document}